\documentclass[aps,prl,twocolumn,superscriptaddress,showpacs]{revtex4}

\usepackage{amsmath}
\usepackage{graphicx}
\usepackage{amssymb}
\usepackage{epstopdf}

\def\ktr{\tilde{\kappa}_{\rm tr}}
\def\kem{\tilde{\kappa}_{e-}}
\def\kep{\tilde{\kappa}_{e+}}
\def\kop{\tilde{\kappa}_{o+}}
\def\kom{\tilde{\kappa}_{o-}}

\def\half{{\textstyle{1\over 2}}}
\def\quar{{\textstyle{1\over 4}}}

\def\EVCR{E_{\rm VCR}}
\def\GVCR{\Gamma_{\rm VCR}}

\def\lsim{\mathrel{\rlap{\lower3pt\hbox{$\sim$}}
    \raise2pt\hbox{$<$}}}
\def\gsim{\mathrel{\rlap{\lower3pt\hbox{$\sim$}}
    \raise2pt\hbox{$>$}}}
\def\sqr#1#2{{\vcenter{\vbox{\hrule height.#2pt
         \hbox{\vrule width.#2pt height#1pt \kern#1pt
         \vrule width.#2pt}
         \hrule height.#2pt}}}}

\def\etal{{\it et al.}}

\newcommand{\beq}[1]{\begin{equation}\label{#1}}
\newcommand{\eeq}{\end{equation}}
\newcommand{\bea}[1]{\begin{eqnarray}\label{#1}}
\newcommand{\eea}{\end{eqnarray}}
\newcommand{\ba}{\begin{array}}
\newcommand{\ea}{\end{array}}

\newcommand{\rf}[1]{(\ref{#1})}

\begin{document}

\title{Particle-accelerator constraints on isotropic modifications of the speed of light}

\author{Michael A.~Hohensee}
\affiliation{Department of Physics, Harvard University, Cambridge, Massachusetts, 02138, USA}
\author{Ralf Lehnert}
\affiliation{Max--Planck--Institut f\"ur Physik,
F\"ohringer Ring 6, 80805 M\"unchen, Germany}
\affiliation{Instituto de Ciencias Nucleares,
Universidad Nacional Aut\'onoma de M\'exico,
A.~Postal 70-543, 04510 M\'exico D.F., Mexico}
\author{David F.~Phillips}
\affiliation{Harvard--Smithsonian Center for Astrophysics, Cambridge, Massachusetts 02138, USA}
\author{Ronald L.~Walsworth}
\affiliation{Department of Physics, Harvard University, Cambridge, Massachusetts, 02138, USA}
\affiliation{Harvard--Smithsonian Center for Astrophysics, Cambridge, Massachusetts 02138, USA}
\date{\today}

\begin{abstract}
The absence of vacuum Cherenkov radiation
for $104.5\;$GeV electrons and positrons at LEP 
combined with the observed stability of $300\;$GeV photons at the Tevatron 
constrains deviations of the speed of light 
relative to the maximal attainable speed of electrons.
Within the Standard-Model Extension (SME), 
the limit 
$-5.8\times~10^{-12}~\leq~\tilde{\kappa}_{\rm tr}-\frac{4}{3}\,c_e^{00}~\leq~1.2\times10^{-11}$ 
is extracted, which sharpens previous bounds 
by more than 3 orders of magnitude. 
The potential for further refinements of this limit
with terrestrial experiments and astrophysical observations 
is discussed.
\end{abstract}



\maketitle

The speed of light, $c$,
has played a crucial role in both 
the conception of Special and General Relativity and its experimental tests. 
The confidence we place in Relativity theory 
is embodied in the fact
that $c\equiv 299\,792\,458\,{\rm m}\!\cdot\!{\rm s}^{-1}$ 
is {\em set} to be a constant 
and provides the basis for the definition of length 
in the International System of Units. 

Currently, 
Relativity tests, 
including precise searches for modifications of the speed of light, 
are experiencing a revival of interest, 
motivated by theoretical studies 
that identify minute violations of Lorentz symmetry 
as a promising imprint of Planck-scale physics~\cite{lotsoftheory}. 
A general theoretical description of 
weak Lorentz symmetry breaking at low energies
is provided by the Standard-Model Extension (SME), 
which contains the usual Standard Model and General Relativity 
as limiting cases~\cite{sme,ssb}. 
To date, 
the SME has served as the basis for numerous Relativity tests 
in many physical systems~\cite{cpt07,kr,GRAAL}.

The majority of potential Relativity violations in electrodynamics 
is governed by the dimensionless $(k_F)^{\alpha\beta\gamma\delta}$ coefficient 
in the SME, 
which causes the speed of light to be direction- and polarization-dependent.
The birefringent components of $k_F$
are bounded down to $10^{-37}$ 
with spectropolarimetric studies of astrophysical sources~\cite{km06}. 
The remaining components 
$\kem$, $\kop$, and $\ktr$ 
are respectively the symmetric, antisymmetric, and trace pieces 
of a $3\times3$ matrix;
they lead to polarization-independent shifts 
of the speed of light.  These parameters 
can be bounded with Michelson--Morley experiments~\cite{her07},
for which effects of $\kem$ are unsuppressed, 
$\kop$ effects are suppressed by $\beta$, 
and $\ktr$ effects are suppressed by $\beta^2$ \cite{secondorderres},
where $\beta\simeq10^{-4}$ is the Earth's orbital speed. 
The corresponding limits are  
$10^{-17}$, $10^{-13}$, and $10^{-8}$, respectively.

These results indicate
that improvements of limits on $\ktr$ 
assume particular urgency. 
Here, we use the analogy 
\beq{analogy}
n=1+\ktr+{\cal O}\left(\ktr^2\right)
\eeq
between $\ktr$ 
and a conventional frequency-independent refractive index $n$
to reduce the sensitivity gap
between $\ktr$ and $\kop$. 
A positive $\ktr$ would imply $n>1$, 
so that the maximal attainable speed (MAS) of other particles
can exceed the speed of light.
This allows charges in a Lorentz-violating ($\ktr>0$) vacuum 
to move faster than the modified speed of light $c/n$ 
and to become unstable against the emission of Cherenkov photons.
A negative $\ktr$ would imply that 
the MAS of charges is now less than the speed of light~\cite{fn1}. 
In this respect, 
the roles of the photon and the charge are reversed 
relative to the $\ktr>0$ case 
suggesting that then the photon is unstable. 
Indeed,
a simple argument shows 
that above some energy threshold 
photon decay into a charge--anticharge pair
is now kinematically allowed.
We employ the absence of these two effects 
for electrons at LEP and photons at the Tevatron
to derive improved limits at the $10^{-11}$ level
on $\ktr-\frac{4}{3}\,c_e^{00}$, 
a quantity that describes differences 
between the speed of light and the electron's MAS.

The quality of such an analysis 
rests on various requirements. 
First, {\em the nature of the charge must be known}:
its MAS serves as the reference 
relative to which the speed of light is constrained~\cite{Hohensee:TBA}.
Second, 
{\em the total rates for vacuum Cherenkov radiation and photon decay 
must be known}. 
Purely kinematical analyses of energy-momentum conservation in these processes are not themselves sufficient if we are to draw conclusions based on their absence~\cite{thres,cher}.
Third, 
{\em the effects of other types of Lorentz violation must be considered}: 
vacuum Cherenkov radiation 
and photon decay could also be generated by, e.g., 
the electron's SME coefficients. 
Note that SME coefficients can typically not be set to zero 
because they may be generated by quantum effects.
A fourth important factor 
is a {\em minimal amount of modeling} 
required to extract the bound.
For example, 
cosmic-ray analyses of Lorentz violation 
necessitate varying degrees of astrophysical 
and shower-development modeling. 

The analysis reported here 
incorporates all four of the above requirements,
providing a 
clean and conservative bound on $\ktr-\frac{4}{3}\,c_e^{00}$.
Our study simultaneously takes advantage of the high-quality data collected at the world's highest-energy accelerators, as well as their superbly controlled laboratory environments.  In so doing, we both improve upon previous constraints, and highlight new avenues for exploring SME physics at existing and future colliders.

Although related tests using observations of ultrahigh-energy cosmic-rays (UHECR) have also sought to constrain Lorentz violating modifications of the fermion--photon vertex~\cite{beall70, cg99,Stecker:2001,jacobson,Altschul:2008,Altschul:2007b,kappatr,ks08}, their conclusions are not directly comparable to our result.  Many UHECR studies do not estimate the rate of vacuum Cherenkov radiation or photon decay, an issue that is nontrivial even for propagation over cosmological distances~\cite{cher}; or they focus on dispersion-relation parameters whose relation to $\ktr$ is unknown or unclear~\cite{beall70,cg99,Stecker:2001,jacobson}.  More recent studies \cite{kappatr,ks08} involve  atomic nuclei as well as electrons, and thus constrain $\ktr-\frac{4}{3}c_{X}^{00}$ for the generally undetermined nucleus $X$ that serves as the UHECR primary scatterer.  Quantitative interpretation of such studies in the broader context of the SME is complicated by both the composite nature of the nuclei as well as the contribution of numerous other SME coefficients that cannot be ignored at UHECR energy scales.  Finally, 
some UHECR investigations of Lorentz-symmetry violation 
require various mild assumptions regarding astrophysical processes~\cite{Altschul:2008,Altschul:2007b}.

The physical system we will consider
consists of photons and electrons,
so we begin by recalling the single-flavor QED limit of the flat-spacetime SME~\cite{sme}:
\bea{lag1}
\mathcal{L} \!& = &\! {}-\quar F^{2}-\quar(k_{F})^{\kappa\lambda\mu\nu}F_{\kappa\lambda}F_{\mu\nu}+(k_{AF})^{\mu}A^{\nu}\tilde{F}_{\mu\nu}\nonumber\\
&&\! {}+\half{\it i}\,\overline{\psi}\,
{\Gamma}^{\nu}
\!\!\stackrel{\;\leftrightarrow}
{D}_{\nu}\! {\psi}
-\overline{\psi}M{\psi}\;,
\eea
where
\bea{defs}
{\Gamma}^{\nu} \! & \equiv & \! {\gamma}^{\nu}+c_e^{\mu \nu}
{\gamma}_{\mu}+d_e^{\mu \nu}{\gamma}_{5}
{\gamma}_{\mu}
\; ,\nonumber\\
M \! & \equiv &  \! m_e
+b_e^{\mu}{\gamma}_{5}
{\gamma}_{\mu}+\half H_e^{\mu \nu}
{\sigma}_{\mu \nu}
\; .
\eea
Here, $F^{\mu\nu}=\partial_{\mu}A_{\nu}-\partial_{\nu}A_{\mu}$ is the
electromagnetic field-strength tensor and
$\tilde{F}^{\mu\nu}=\tfrac{1}{2}\epsilon^{\mu\nu\rho\sigma}F_{\rho\sigma}$
denotes its dual.  
The spinor $\psi$ describes electrons of mass $m_e$, 
and the usual U(1)-covariant derivative 
is denoted by $D^\mu=\partial^\mu+i e A^\mu$. 
The spacetime-independent SME coefficients 
$(k_{F})^{\mu\nu\rho\lambda}$, $(k_{AF})^{\mu}$, 
$b^\mu$, $c^{\mu\nu}$, $d^{\mu\nu}$, and $H^{\mu\nu}$
control the extent of Lorentz and CPT violation.

In what follows, we are primarily interested in the
$\ktr$ component of the CPT-even $(k_{F})^{\mu\nu\rho\lambda}$. 
The $k_{F}$ coefficient exhibits the symmetries of the Riemann tensor, 
and its double trace $(k_{F})^{\mu\nu}{}_{\mu\nu}$ vanishes.
This leaves 19 independent components. 
In a given coordinate system, 
which is conventionally chosen to be the Sun-centered celestial equatorial frame, 
$k_{F}$ can be decomposed as follows~\cite{km0102}.
Ten components 
are associated with birefringence 
and can be grouped into the two dimensionless $3\!\times\!3$ matrices 
$\kom$ and $\kep$. 
The remaining nine components 
\beq{ktildedef}
\tilde{k}^{\mu\nu}\equiv (k_F)_\alpha{}^{\mu\alpha\nu}
\eeq
give rise to $\kem$, $\kop$, and $\ktr$, 
as explained earlier. 
In particular, 
$\ktr=-\tfrac{2}{3}(k_{F})^{0j0j}$, 
where the index $j$ runs from 1 to 3 
and is summed over in this expression. 

All of the SME coefficients in Lagragian~\rf{lag1} 
modify either the photon's or the electron's dispersion relation 
and therefore also the kinematics of the electron--photon vertex. 
It follows 
that $\ktr$ cannot be singled out
in studies of vacuum Cherenkov radiation and photon decay;
the other SME coefficients for the photon and the electron 
must in general also be taken into account.
However, 
a careful analysis of previous experiments reveals
existing stringent limits on 
these additional SME parameters~\cite{Hohensee:TBA}.
The scale of these limits 
is governed by
\begin{equation}
\mathcal{S}\equiv \max{\left(\kem,\kep,\kop,\kom,\frac{k_{AF}}{m_{e}},\frac{b}{m_{e}},c,d,\frac{H}{m_{e}}\right)}\label{scale}\;,
\end{equation}
where the absolute values of the individual components listed here are implied.
At present, $\mathcal{S}\sim 10^{-13}$ is dominated by $\kop$ \cite{kr}.
This value is more than five orders of magnitude smaller than
present limits on $\ktr$, 
and about a factor of $10^2$ smaller than 
the bound on $\ktr$ 
we derive here.
We therefore can safely ignore other SME coefficients in our analysis 
and retain $\ktr$ only. 

A related issue concerns the physical equivalence 
of the photon's $\tilde{k}^{\mu\nu}$ and electron's $c_e^{\mu\nu}$.
These two coefficients 
are associated with the same phenomenology, 
and they can therefore not be distinguished 
within the framework of Lagragian~\rf{lag1}. 
In the present context, 
our $\ktr$ model is physically equivalent 
to a model with 
\begin{align}
\label{equiv}
c_e^{00}&=-\tfrac{3}{4}\,\ktr &\text{and } &&c_e^{jj}&=-\tfrac{1}{4}\,\ktr\;.
\end{align}
In this expression,
there is no sum over $j=1,2,3$.
This equivalence 
can be established rigorously 
with a coordinate rescaling~\cite{km0102,bk04},
which implies that only $\ktr-\frac{4}{3}\,c_e^{00}$ 
is observable in the context of Lagrangian~\rf{lag1}.
This rescaling can be used 
to remove either $\ktr$ or $c_e^{00}$ 
from the model. 
We often select coordinates such that $c_{e}^{00}=0$, 
but undo this special rescaling and reinstate $c_{e}^{00}$ when quoting results.

At leading order, 
the photon dispersion relation 
in the presence of $\ktr$ is given by \cite{km0102}
\begin{equation}
E_{\gamma}^2-(1-\ktr)\vec{p}\!\phantom{.}^{2}=0\;,\label{eq:dispersion}
\end{equation}
where $p^{\mu}\equiv (E_{\gamma},\vec{p})$ is the photon's 4-momentum. 
Thus, 
the speed of light is $(1-\ktr)$, and in the present context, in which we treat the fermion dispersion relation as being unaffected by Lorentz violation, vacuum Cherenkov radiation can only occur for positive $\ktr$.  
Using Eq.~\rf{eq:dispersion} 
and energy--momentum conservation 
for the emission of a Cherenkov photon
yields the energy threshold~\cite{Altschul:2008}
\begin{equation}
E_{\rm VCR}=\frac{1-\ktr}{\sqrt{(2-\ktr)\ktr}}\,m_e=\frac{m_e}{\sqrt{2\ktr}}+\mathcal{O}\left(\sqrt{\ktr}\right).\label{vcrenergy}
\end{equation}
For electrons with energies above $E_{\rm VCR}$, 
vacuum Cherenkov radiation is kinematically allowed. 
Equation~\rf{vcrenergy} 
can alternatively be derived from 
the usual Cherenkov condition 
that the electron must be faster 
than the speed of light $(1-\ktr)$. 

We extract a limit on $\ktr$ through 
the absence of observed vacuum Cherenkov radiation.  
This requires the emission to be efficient enough
that charges with energies above $\EVCR$ 
are rapidly decelerated below threshold.  
Near $\EVCR$, 
the dominant deceleration process is
single-photon emission 
with the estimated rate~\cite{Altschul:2008}
\begin{equation}
\GVCR=\alpha\, m_e^{2}\,\frac{(E_{e}-\EVCR)^{2}}{2E_{e}^{3}}\;,\label{vcrate}
\end{equation}
where $\alpha$ is the fine-structure constant,
and $E_{e}$ denotes the electron energy.  
This expression shows 
that the emission process is quite efficient, 
and we now use it 
to derive limits on $\ktr$ 
from the energies 
attained by primary electrons at the LEP collider.  
The highest laboratory-frame particle energy reached at LEP was
$E_{\rm LEP}=104.5$ GeV with a relative uncertainty in the center-of-mass energy $\Delta E_{\rm CM}/E_{\rm CM}$ below $2.0\times 10^{-4}$ \cite{LEP05}.  Using Eq.~\eqref{vcrate}, we find that if $\EVCR=104\;$GeV, electrons initially accelerated to $104.5\;$GeV would be rapidly slowed by emission of Cherenkov photons to an energy below $\EVCR$ over a $1/e$ length of about $95\;$cm.  The total energy lost to the Cherenkov effect in such a scenario would far exceed the value allowed by measurements~\cite{Hohensee:TBA}.  With Eq.~\eqref{vcrenergy}, 
  the requirement that $\EVCR$ be greater than $104\;$GeV becomes
%
%
\begin{equation}
\ktr-\tfrac{4}{3}\,c_e^{00} \leq 1.2\times 10^{-11}\;,\label{VCRbound}
\end{equation}
where we include the dependence on $c_{e}^{00}$.
%
This bound is significantly smaller than previous laboratory limits on $\ktr$. 
Note also that 
the scale $\mathcal{S}$ defined in Eq.~\eqref{scale} 
is not yet reached,
which justifies the exclusion of 
other photon or electron SME coefficients 
in our analysis.

For negative $\ktr<0$,
the dispersion relation~\eqref{eq:dispersion} remains valid,
and photons may travel faster than 
the MAS of electrons~\cite{fn1}.  
This precludes vacuum Cherenkov radiation 
at the cost of eliminating photon stability: 
for $E_\gamma$ above the threshold 
\begin{equation}
E_{\rm pair}=\frac{2m_e}{\sqrt{\ktr(\ktr-2)}}=\sqrt{\frac{2}{-\ktr}}m_e
+\mathcal{O}\left(\sqrt{\ktr}\right)\; ,\label{gammathres}
\end{equation}
photon decay into an electron--positron pair 
is kinematically allowed~\cite{cg99,Hohensee:TBA}.
The corresponding leading-order decay rate
is given by~\cite{ks08,Hohensee:TBA}
\begin{equation}
\Gamma_{\rm pair}=\frac{2}{3}\,\alpha\, E_{\gamma}\,\frac{m_e^{2}}{E_{\rm pair}^{2}}\,
\sqrt{1-\frac{E_{\rm pair}^{2}}{E_{\gamma}^{2}}}\left(2+\frac{E_{\rm pair}^{2}}{E_{\gamma}^{2}}\right)\;.\label{pairrate}
\end{equation}
Paralleling the Cherenkov case, 
this process is also highly efficient.  
For example, 
a $40\,$GeV photon with energy 1\% above threshold 
would decay after traveling
about $30\, \mu$m.

The above reasoning establishes 
that the existence of long-lived photons 
with high energies 
constrains negative values of $\ktr$.  
Photons generated in terrestrial laboratories 
provide a clean, well-characterized source for bounding $\ktr$.  
Although the accessible energies 
are lower than those in cosmic rays,
terrestrial tests offer larger data samples 
and a better control of the experimental conditions. 
Hadron colliders produce the highest-energy photons 
and therefore yield tight Earth-based experimental limits on $\ktr$.  
Thus, we consider Fermilab's Tevatron $p\overline{p}$ collider 
with center-of-mass energies up to $1.96\,$TeV.
Isolated-photon production with an associated jet 
is important to QCD studies 
and has been investigated with the D0 detector. 
The recorded photon spectrum 
extended up to a single event at $442\,$GeV~\cite{D006},
which would imply $\ktr\gsim-3\times 10^{-13}$.
While such a single event 
is insufficient to draw conservative conclusions regarding Lorentz symmetry, 
it is indicative of the sensitivity of this method.

We restrict our analysis to lower-energy D0 photon data with good statistics, 
where photons with energies up to $340.5\,$GeV were observed~\cite{D008}, 
and comparisons to QCD theory were made.  
The $340.5\,$GeV bin extended from $300\,$GeV to $400\,$GeV; 
the measured flux deviated 
by a factor of $0.52\pm0.26$ 
from QCD predictions~\cite{D008}, 
so that at most $74\%$ 
of the produced photon flux 
can be lost due to hypothetical photon decay.
We continue by conservatively assuming 
that all events in this bin were $300\,$GeV photons, 
and we take  $E_{\rm pair}=300\,$GeV. 
This is again justified by the rapid photon-decay rate~\rf{pairrate}:
if $E_{\rm pair}$ were just $0.1\,$keV 
below the lowest observed $300\,$GeV photon energy, 
the photon deficit would be larger 
than the allowed $74\%$~\cite{Hohensee:TBA}. 
In other words, 
the uncertainty in $E_{\rm pair}$ 
is essentially determined by 
the resolution of the photon-energy measurement.
This reasoning gives the limit
\begin{equation} 
-5.8\times10^{-12}\leq\ktr-\tfrac{4}{3}\,c_e^{00}\;, 
\label{decaybound}
\end{equation}
where we again include the contribution of $c_e^{00}$.
Like the Cherenkov constraint, 
the photon-stability limit~\rf{decaybound} is larger than the scale ${\cal S}$, 
so other photon- or electron-sector coefficients 
are not further constrained by this argument.  
At the same time, 
this justifies the exclusion of these additional coefficients 
from our study.

Combining the bounds~\rf{VCRbound} and~\rf{decaybound}, 
we obtain the two-sided limit
\begin{equation}
-5.8\times10^{-12}\leq\ktr-\tfrac{4}{3}\,c_e^{00}\leq1.2\times10^{-11}
\label{bound}
\end{equation}
on isotropic deviations of 
the phase speed of light from 
the MAS of the electron.
This bound represents an improvement of more than 3 orders of magnitude
upon previous laboratory-based constraints.
We obtained this limit by 
exploiting the threshold effects of
vacuum Cherenkov radiation
and photon decay
for positive and negative $\ktr$,
respectively.

An independent constraint on $\ktr$ 
may be obtained by 
future low-energy laboratory tests 
with estimated sensitivities at the level of $10^{-11}$ or better~\cite{future}.  
Another possibility for improvents may come from photon triple splitting,
because the amplitude for this effect is nonzero 
in the presence of $c^{\mu\nu}$-type SME coefficients~\cite{triple}.  
As opposed to vacuum Cherenkov radiation and photon decay, 
photon triple splitting is not a threshold effect, 
so that it may not necessitate high photon energies.

Significantly improved terrestrial bounds 
using the same reasoning as presented here 
may be obtained at the prospective International Linear Collider (ILC).  
Accelerating electrons to laboratory-frame energies of $500\,$GeV, 
the ILC may provide a one-sided Cherenkov bound with a sensitivity 
at the level of $5\times10^{-13}$.  Similarly, the Large Hadron Collider (LHC) 
is scheduled to attain roughly seven times the energy of the Tevatron.
Assuming that the energy $E_{\gamma}$ of the produced photons scales by the same factor, 
the bound of Eq.~\eqref{decaybound} can be sharpened by 
a factor of 50.  
Additional improvements of the photon-decay limit 
may be achieved with a dedicated D0 or LHC study: for instance,
the
highest-energy data not analyzed for QCD tests 
could be used for the present purposes.
Moreover, 
the high-energy tail of the photon-energy spectrum
could be utilized more efficiently by 
avoiding large energy binning.

The $\ktr$ limits from both 
vacuum Cherenkov radiation and photon decay 
scale quadratically with the energy of the primary particle.
At present,
UHECR offer the highest possible energies 
in the Sun-centered celestial equatorial frame; 
the spectrum is limited only by
the opacity of the universe to cosmic rays above certain
energy thresholds 
(e.g., GZK suppression or pair creation on IR-photon background).
For example, 
particles with energies up to about $2\times10^{11}\,$GeV
have been observed at the Pierre Auger Observatory. 
Assuming 
these particles are Iron nuclei, and that the neutron $c_n^{\mu\nu}$ coefficients are insignificant,
bounds at the $10^{-20}$ level 
can be extracted~\cite{ks08}. 
Although this limit is not directly comparable to our results 
(it does not measure the photon speed
relative to the MAS of the electron), 
it does illustrate the potential of cosmic-ray tests.
Primary photons from the Crab nebula are another example: 
Energies up to $8\times10^{4}\,$GeV 
have been reported by HEGRA~\cite{HEGRA04}.
Equation~\rf{gammathres} would then give one-sided
limits on $\tilde{\kappa}$ coefficients at the level of $10^{-16}$.
We mention that 
at UHECR scales 
some of the non-birefringent $\tilde{\kappa}$ matrices
and certain electron SME coefficients can no longer be neglected, 
as is apparent by consulting Eq.~\eqref{scale}.  In any case, 
the interpretation of future UHECR analyses of Lorentz violation
would greatly benefit from 
a more reliable identification of the primary.

\acknowledgments 
The authors would like to thank B.~Altschul for his helpful comments as this work developed, as well as 
 F.R.~Klinkhamer and M.~Schreck for useful discussions.
This work is supported in part by the National Science
Foundation and by the European Commission
under Grant No.\ MOIF-CT-2005-008687.


\begin{thebibliography}{15}
\bibitem{fn1}
This does not necessarily imply causality violations~\cite{caus}:
we can rescale the coordinates
such that the new lightcone coincides 
with the fastest particle~\cite{km0102,bk04}.
\end{thebibliography}
\end{document}